# An AI-directed analytical study on the optical transmission microscopic images of Pseudomonas aeruginosa in planktonic and biofilm states

Bidisha Sengupta [1,4,*], Mousa Alrubayan [2,4], Yibin Wang [3,4], Esther Mallet [1], Angel Torres [1], Ravyn Solis [1], Haifeng Wang [3,*], Prabhakar Pradhan [2,*]

[1]Department of Chemistry & Biochemistry, Stephen F. Austin State University, Nacogdoches, TX, 75962
[2]Department of Physics and Astronomy, Mississippi State University, Mississippi State, MS 39762.
[3]Department of Industrial Engineering, Mississippi State University, Mississippi State, MS 39762.
[4]These authors have equal contributions.
[*]Corresponding Authors: BDasgupta, HWang, and PPradhan

**Abstract:** Biofilms are resistant microbial cell aggregates that pose risks to health and food industries and produce environmental contamination. Accurate and efficient detection and prevention of biofilms are challenging and demand interdisciplinary approaches. This multidisciplinary research reports the application of a deep learning-based artificial intelligence (AI) model for detecting biofilms produced by *Pseudomonas aeruginosa* with high accuracy. Aptamer DNA templated silver nanocluster (Ag-NC) was used to prevent biofilm formation, which produced images of the planktonic states of the bacteria. Large-volume bright field images of bacterial biofilms were used to design the AI model. In particular, we used U-Net with ResNet encoder enhancement to segment biofilm images for AI analysis. Different degrees of biofilm structures can be efficiently detected using ResNet18 and ResNet34 backbones. The potential applications of this technique are also discussed.

## 1. Introduction

Biofilm-forming microbes synthesize different signaling biomolecules such as proteins, carbohydrates, and DNA, which help in their cooperative activities and attachments to biotic or abiotic surfaces to create organized multicellular communities called biofilm [1–3]. Studies on biofilm by the *gram*-positive bacterium were pioneered by Ferdinand Cohn in 1877 [1]. Numerous studies on biofilms of various *Bacillus* species (*Bacillus cereus, Bacillus anthracis,* and *Bacillus thuringiensis, BT*) [4–7] and other bacterial species (*gram*-negative *Pseudomonas aeruginosa, PA*) were conducted in the last decade to understand the mechanism of the growth and control of biofilms [8,9]. Many medically important fungi form biofilms [10–12] (including yeast *Candida* [13] and mold [10], which are resistant to antifungal drugs. Biofilm formation causes chronic infections and loss of immune responses in individuals with underlying health problems and can even lead to death, creating a severe danger to public health [14–16]. This is due to the resistance of the biofilms against antimicrobial agents [14]. Biofilms also cause significant problems by contaminating medical devices, food, and the environment [17,18], threatening healthcare industries, including hospital devices, human health, and biotech [19,20] companies. According to a study by the NIH [21], biofilm formation can be attributed to a significant portion (65% and 80%) of all microbial and related recurring infections.

Designing effective strategies to combat biofilm-related infections is crucial, as it demands a detailed understanding of the underlying processes in biofilm formation and unraveling their scientific implications. Phenotypes of biofilm-forming microbes differ from their planktonic states, leading to the rise of antibiotic resistance and the failure of treatment of biofilm infection by modern antibiotics [14,22]. An alternative approach to combat these deadly biofilms needs to be addressed. Nanotechnology has become prominent in reducing and controlling biofilm formation [23–26]. In personal and medical care and household products, silver is one of the most widely used metals [27–29]. Recently, Ag nanoparticles (NP) were used to prevent biofilm in *gram*-positive and *gram*-negative bacteria [28,30–32]. Small DNA stretches are designed to make aptamer molecular beacons, called functionalized aptamers [33–35], for detecting biofilms. Sengupta, B. group has pioneered the usefulness of DNA aptamer to scaffold and carry silver nanoclusters (Ag-NC) to influence biofilm formation [7,8]. They have also shown that a capping agent like methyl-beta-cyclodextrin (CDx) [7] bound Ag-NC had a better preventive effect on biofilm.

Recently, AI approaches such as convolutional neural networks (CNN) have come into prominence for biofilm (object) detection [36–39]. Effective AI model Unet [37] is widely recognized as suitable and has the potential for different types of biomedical image segmentation. The network architecture contains a fully convolutional encoder-decoder with skip connections between the encoder blocks and their symmetric decoder blocks. ResNet [37,40] introduces the concept of skip connections, allowing the network to skip specific layers to let the model learn effectively. It can be shown that a U-Net framework built with a ResNet encoder can leverage the power of deep residual learning and enhance the feature extraction process to improve the overall segmentation performance [41]. Identification of biofilm is difficult for humans to identify with visual information; extensive and essential features related to biofilm are expected to be captured through a deep network learning process. Biofilm bright field images can be characterized using U-Net architecture with ResNet18 and ResNet34 backbone. This research exploits an AI-based model to predict and detect the microbial biofilm of *gram*-negative *Pseudomonas aeruginosa,* PA [8] with higher accuracy against a given aptamer-DNA-based Ag-NC. Our results show that this AI model can be applied to any image to detect biofilm formation with a higher accuracy.

In this paper, we report the prevention of PA biofilm by the Ag-NC, which is synthesized on a DNA aptamer matrix. In particular, we created biofilm using PA in 2D fully or partially on a glass slide, along with or without the treatment of PA targeted DNA aptamer enclosed Ag-NC. Large-volume bright-felid images of the PA biofilms grown on glass slides were then generated using an automated transmission microscope. Finally, we analyze the large-volume bright field data via AI, especially using U-net with ResNet. For the AI analyses, the AI algorithm is first fed with bright-field images with marked biofilms, then unknown biofilms are fed to the leaned AI model for detection. Results show the detection of biofilm formation/prevention with higher accuracy.

**2. Experimental Methodology and Instrumentation**

*2.1. Preparation of DNA-templated Silver Nanocluster*

The silver nanocluster (Ag-NC) was prepared on the aptamer DNA 5'-CCC CCG TTG CTT TCG CTT TTC CTT TCG CTT TTG TTC GTT TCG TCC CTG CTT CCT TTC TTG-3' (which is specific for PA [42]) following the protocol published elsewhere [43]. This DNA oligonucleotide was custom-synthesized from Integrated (DNA) Technologies (IDT, USA). Lyophilized DNA was hydrated with triple distilled water (obtained from Sigma). Ag-NC was synthesized by combining 15 µM DNA and 90 µM $Ag^+$ with 6 $BH_4^-$ /oligonucleotide solutions, followed by vortex mixing for 1 min. The sample was stored overnight in the dark at 4°C. UV/V is absorption, and fluorescence emission studies were conducted in the solution.

*2.2. Preparation of Bacterial Samples for Biofilm Study*

*Pseudomonas aeruginosa* ATCC 10145, Lot 416-116-4, was obtained from Microbiologics Inc, MN. PA was grown in 200 mL Tryptic Soy Broth (TSB, FisherSci.) at 23°C overnight. 25% of Ag-NC solution was added to media containing bacterial culture. In a separate study, PA in 100% media and in the presence of 25% water (the same volume added for Ag-NC) showed little difference in bacterial growth. Hence, this work has used PA culture with 25% water as a control. Plates were incubated at 23°C for two days in six-well plates. Following our previous work, turbidity in the wells indicated biofilm formation [7,44]. Cells from the plates were heat-fixed on the glass slides for bright-field imaging.

*2.3. Steady State Absorption, Fluorescence*

We performed steady-state absorption and fluorescence spectroscopic measurements to confirm Ag-NC formations. Steady-state absorption spectra were recorded with a Shimadzu UV 2550 spectrophotometer. Steady-state fluorescence measurements were carried out with a PerkinElmer FL 6500 fluorescence spectrophotometer. Excitation and emission slit widths were 5/10 nm. All reported luminescence spectra were corrected for the detector's spectral response.

The fluorescence and absorption experiments were performed to characterize spectral properties of the DNA aptamer scaffolded silver nanoclusters, which were reported in our previous work [8]. We repeated these measurements to ensure the reproducibility of the Ag-NC formation on the DNA-templated aptamer and were able to reproduce almost similar results as reported in [8]. The repeated measured fluorescence and absorption spectra have been reported in Appendix A for the completeness of this article. In brief, for example, fluorescence emission spectra of aptamer-DNA 5'-CCC CCG TTG CTT TCG CTT TTC CTT TCG CTT TTG TTC GTT TCG TCC CTG CTT CCT TTC TTG-3' templated Ag-NC show the absorption band around the wavelength region 380-390 nm. An absorption band peaking around 427 nm, with a shorter band peaking around ~530 nm, was observed. This indicates the formation of more than one type of silver nanoclusters. Furthermore, the emission spectra of Ag-NCs showed stronger fluorescence with $\lambda_{em}^{max}$ at ~ 633 nm for $\lambda_{ex}$ = 540 nm, compared to the emission band with $\lambda_{em}^{max}$ at 530 nm for $\lambda_{ex}$ = 450 nm, agreeing with our earlier work [8].

*2.4. Instrumentation*

Olympus BX61 optical microscope, CCD camera, and PRIOR Test Control (OptiScan software) were used for large-volume bright-field transmission imaging. Rather than the traditional way of manual imaging, we used the auto-scan feature of the microscope to identify and diagnose the biofilm prevention/formation and their statistical mixtures in thin biofilms on glass slides. To image the biofilms, we used a conventional Olympus BX61 (Figure 1) motorized system microscope with a 40x objective (UIS2) series, and a CCD camera mounted on the top of a BX61 microscope. In addition, we used the PRIOR Test Control via OptiScan software to move the microscope stage to synchronize with the camera while taking the scattering micrographs in the transmission mode.

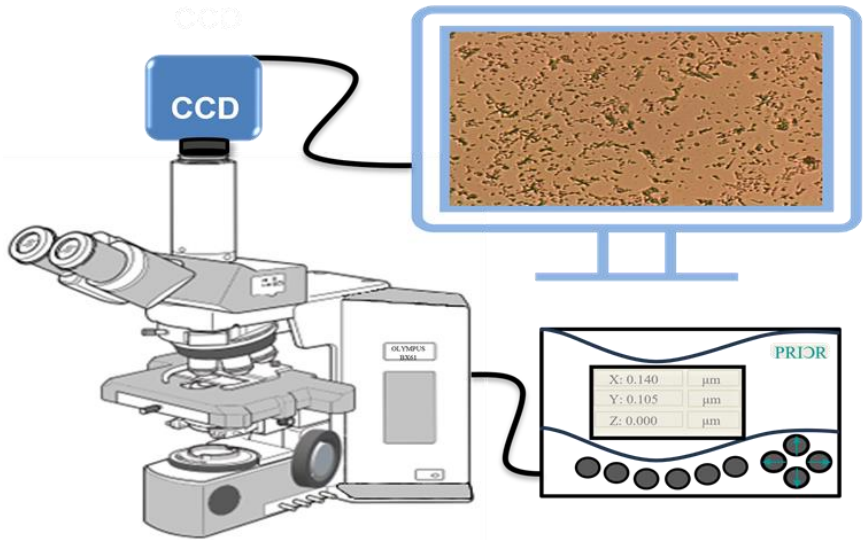

**Figure 1.** Schematic picture of a large volume bright-field imaging system using Olympus Bx61 microscope (this picture is based on Olympus manual guide) and Prior automated control system. The slide was kept on the slide holder on the x-y-z automated stage of the microscope and then scanned by a programmable matrix array. The scan speed is around ~1000 spot /hr.

*2.5. Scanning Method*

The most significant squares within each observation site of the biofilm over glass slide were determined and divided into numbers proportional to the objective lens used (40x). Altogether, ~ 2000 microscopic micrographs were generated from each type of sample (C, CN1, CN2, CN3…) for the AI analyses.

*2.6 AI: ResNet-based U-Net Biofilm Segmentation*

In this section, a U-Net-based segmentation model is used to segment biofilm to further retrieve the demanded quantitative measures of the bacteria, such as the covered area and cell count. We performed a pixel-level biofilm semantic segmentation using a U-Net1-based [40,45] AI framework within the studied regions. The analyzed dataset includes a total of 184 microbial images with annotated biofilm masks. Data robustness is ensured by tolerating variations in its collection and annotation procedure, where one image sample may contain from none to multiple segmented biofilm regions. The dataset is divided into a training set with ~150 samples and a test set with 34 samples. Each image data with annotated masks is resized to the same input shape, 512×512. The pixel values are normalized between 0 and 1. A preprocessed image sample with its ground truth mask is shown in Figures 2A and 2B, respectively. The bright-field imaging system captured the input image data, and domain experts collected and validated the corresponding manually annotated masks (yellow areas indicate biofilm).

UNet is widely recognized as an effective AI model for biomedical image segmentation. The network architecture contains a fully convolutional encoder-decoder with skip connections between the encoder blocks and their symmetric decoder blocks. This encoder-decoder structure of U-Net has inspired many segmentations mean in medical imaging; for instance, the attention mechanism has been employed in medical image segmentation and become widely adopted. The variation of U-Net-related deep learning networks is designed to optimize results by improving medical image segmentation's accuracy and computing efficiency through changing network structures.

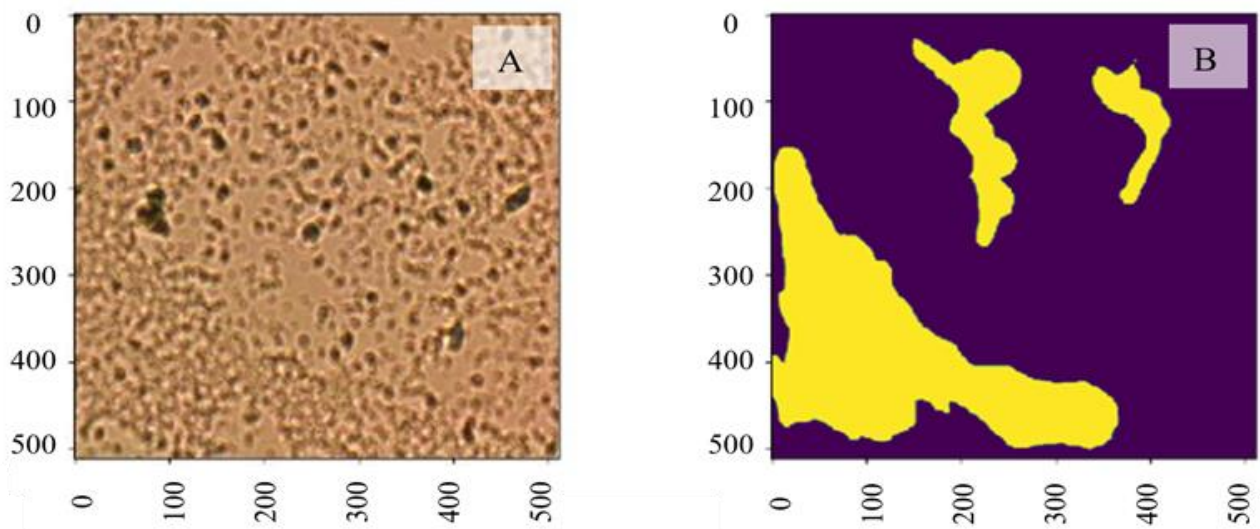

**Figure 2**: A. Input image data captured by the bright-field imaging system; B. The corresponding manually annotated mask (yellow areas indicate biofilm).

The convolution scheme is modified and extended in the conventional U-Net framework to work with few training images and produce more accurate segmentation. The general shrinkage network is replaced with sequential layers. The high resolution of the contracted path is combined with the upsampled output for localization. Therefore, sequential convolutional layers can study informative features and output more accurate segmentation. The network applies the practical part of every convolution, where the segmentation map contains mere pixels, and the complete context of the pixels can be obtained in the input image.

ResNet [37–39] introduces skip connections, allowing the network to skip specific layers to let the model learn effectively. A U-Net framework built with a ResNet encoder can leverage the power of deep residual learning and enhance the feature extraction process to improve the overall segmentation performance [41]. In addition, the shortcut mechanism added by the ResNet tends to avoid gradient vanishing and improve the network convergence efficiency. Because biofilm is even more complex for humans to identify with visual information, extensive and essential features related to biofilm are expected to be captured through a deep network learning process. The proposed ResNet-based U-Net segmentation framework is shown in Figure 3. This study investigated U-Net architecture with ResNet18 and ResNet34 backbone, respectively.

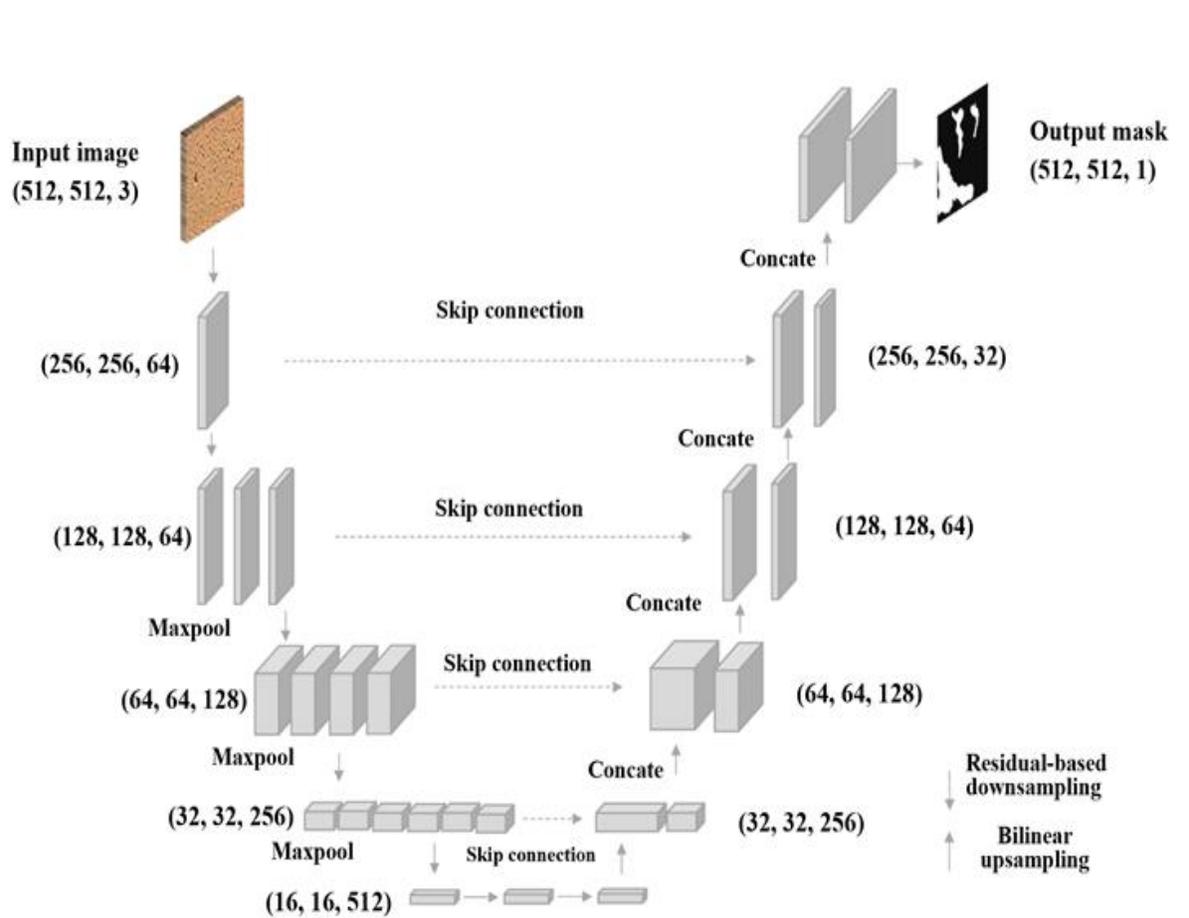

**Figure 3**: Architecture of the U-Net model. The model takes the input with an image size of 512×512×3. The downsampling process performs the feature extraction and compression, and the upsampling performs segmentation map generation. The output of the model is the biofilm probability map.

## 3. Results and Discussions

Figure 1 shows representative bright field images of control (Figure 4A) and aptamer-enclosed nanocluster-treated (Figure 4B) samples of *Pseudomonas aeruginosa.* Incubation of the PA cells with the aptamer-enclosed Ag-NC for 48 hours cleared the turbidity (see Figure 4B inset), which was observed in control and was indicative of biofilm formation [7,8] (as shown in Figure 4A).

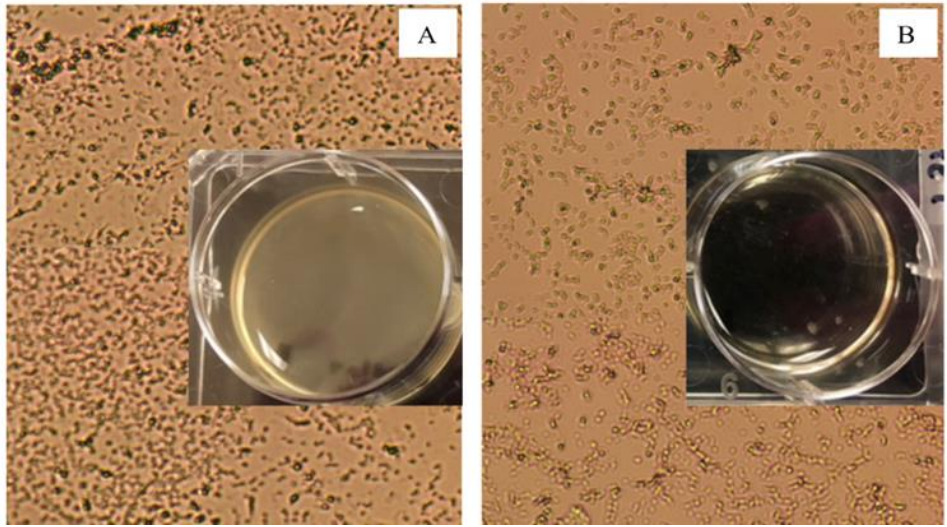

**Figure 4**: Brightfield images of *Pseudomonas aeruginosa* (PA) in Tryptic Soy Broth culture medium in the absence (control (A)) and presence of aptamer-DNA templated silver nanoclusters (B). The biofilm formation is evident by the turbidity in the control state (1A inset), while planktonic cells are observed in the Ag-NC sample (1B inset). Turbidity in well A containing control PA solution proves biofilm formation, which is cleared in the presence of Ag-NC (well B). Both the wells were photographed on a black background.

*Machine Learning Results for Biofilm Segmentation*

The AI frameworks' training inputs are image data with their corresponding biofilm annotations, and the output is the model-predicted biofilm mask. The binary cross entropy loss between the actual label and the prediction is computed and minimized in the training. The Adam optimizer is determined with a learning rate of 5e-4. The ResNet18 and ResNet34 backbones use a batch size of 16, respectively.

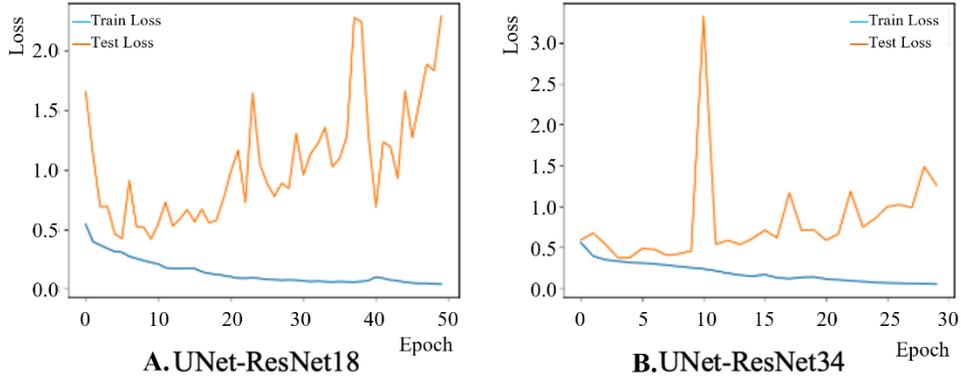

**Figure 5**: A comparison of training loss between the segmentation model with different backbone frameworks A. U-Net-ResNet18 and B. U-Net-ResNet34.

The training loss between different backbone learning frameworks is shown in Figure 5. The models started overfitting after about ten epochs as seen in Figures 5A and 5B, which is the point where training loss keeps decreasing but test loss starts to increase. Compared to the ResNet18-based backbone, the U-Net framework with ResNet34 possessed a smoother training trend. The best-performed model was saved for the test. We report the model test accuracy, precision, recall, F-1 score, and IoU (Intersection over Union) in Table 1. The results show that ResNet-34-based U-Net architecture achieved a better segmentation test performance with 86.62% accuracy, 71.12% precision, 73.84% recall, 71.43% F-1 score, and 59.70% IoU. It is also demonstrated that more informative biofilm features are learned through a deeper encoder architecture.

**Table 1.** Test performance comparison of the studied AI models

| Model | Accuracy | Precision | Recall | F-1 Score | IoU |
|---|---|---|---|---|---|
| U-Net-ResNet18 | 0.8469 | 0.6322 | 0.6948 | 0.6554 | 0.5360 |
| U-Net-ResNet34 | 0.8662 | 0.7112 | 0.7384 | 0.7143 | 0.5970 |

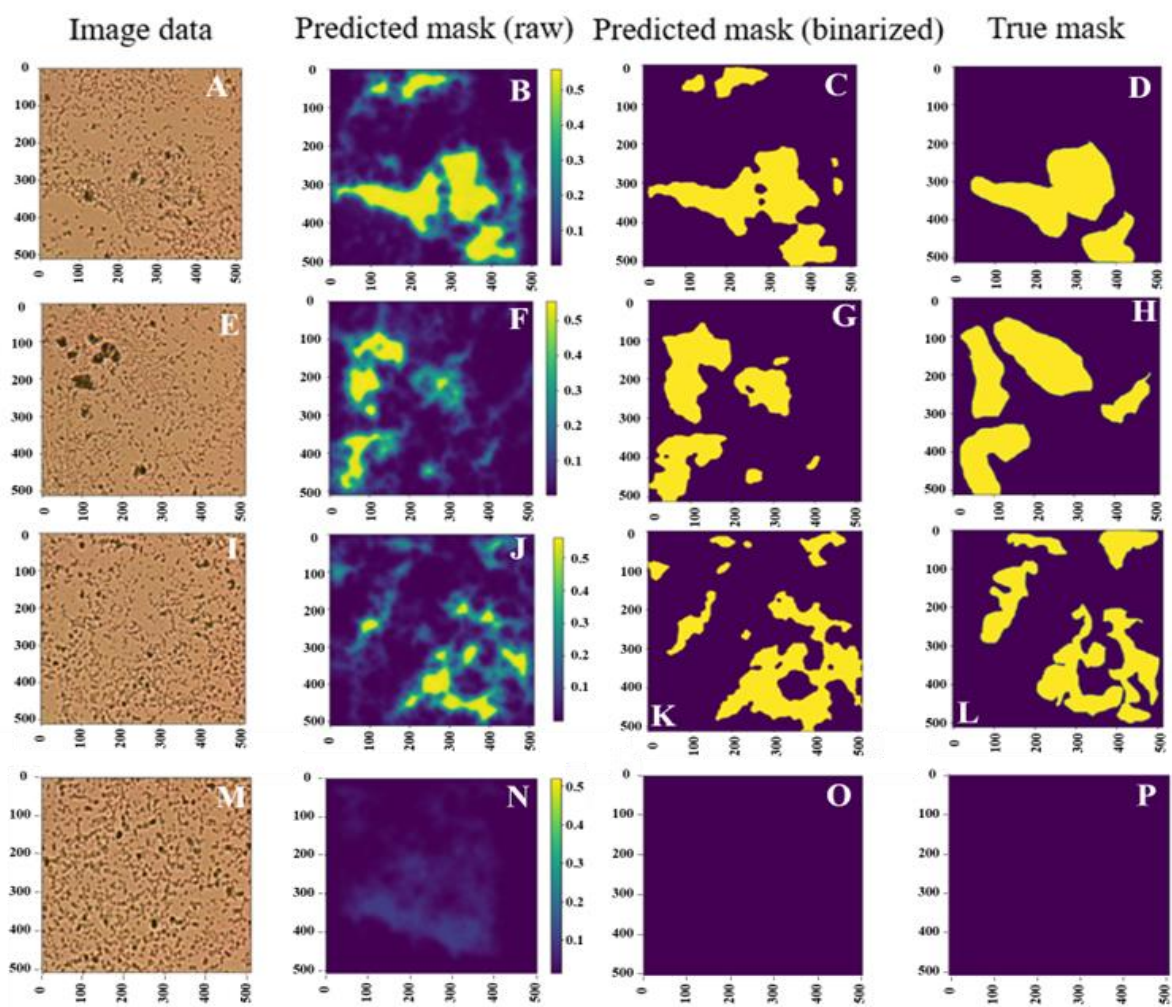

**Figure 6:** A test comparison of segmentation model raw output, predicted mask, and ground truth mask for different micrographs. Based on the image input (1st column), the U-Net-ResNet34 algorithm generates the biofilm probability map (2nd column). Otsu thresholding is applied to binarize the biofilm probability map to generate the predicted mask (3rd column). The fourth column shows the true mask provided by human expert as the comparison with the 3rd column.

Post-processing was performed to generate binary predicted masks based on the output predicted probability maps. Mainly, Otsu thresholding [46] was applied to binarize the segmentation prediction. A comparison of the model raw output indicated mask and ground truth mask is shown In Figure 3; the first column has three different biofilm samples of *Pseudomonas aeruginosa* (PA) (Figures 6A, 6E, and

6I). Figures 6D, 6H, 6L, and 6P illustrate the related true masks provided by human experts. Note that Fig. 6M is an image of *Pseudomonas aeruginosa* (PA) cells in a planktonic state. Based on each image input, the U-Net-ResNet34 algorithm, as shown in Figure 6, generates the related biofilm probability map as shown in Figures 6B, 6F, 6J, and 6N, where yellow shows the probability of being biofilm. Otsu thresholding is applied to binarize the biofilm probability map (Figures 6B, 6F, 6J, and 6N) to generate the predicted mask, as shown in Figures 6C, 6G, 6K, and 6O. The outperformed U-Net-ResNet34 architecture predicts the masks. It is demonstrated that the expected biofilm regions are aligned with the ground truth annotations, and the general shape of the biofilm is indicated. Hence, the ResNet-based U-Net method has great potential in this studied biofilm segmentation task. The experimental results demonstrated that the proposed model could achieve effective segmentation performance by generating accurate biofilm predictions compared to ground truth masks. Further quantitative measures can be computed and analyzed with this prediction.

## 4. Conclusions

We experimented to prevent biofilm formation using silver nanoclusters, which were synthesized on DNA aptamer matrices and were used to treat the bacterial biofilms of *Pseudomonas aeruginosa*. The experimental results showed that in the presence of Ag-NC, the degree of the 2D biofilm formation decreased. The biofilm formation/prevention statistical analysis was performed using AI-based ResNet 18 and ResNet34. In particular, the AI algorithm was trained for two cases: the formation and prevention of biofilm, followed by testing of the AI model for its ability to detect biofilm among many images of *Pseudomonas aeruginosa*. The results show the accuracy of detection is 85-87%. The developed technique can be used for the fast detection of biofilms by healthcare, biotech and environmental agencies.

**Author Contributions:** PP, BS, and HW conceived the idea; BS developed the biofilms; MA performed the large-volume brightfield imaging of the biofilms; EM, AT, and RS identified the biofilm areas on the images; MA, YW, and HW performed the AI analyses; PP, BS, HW, and MA wrote the first draft of the paper; all authors contributed to the final version of the paper; and PP monitored the overall project.

**Funding:** PP acknowledges partial support from the National Institute of Health (Grant No. R21CA260147); HW acknowledges support from the National Institute of Health (Grant No. R03DE032766). BS acknowledges support from the Welch Foundation Grant (No. AN-0008) at the Department of Chemistry and Biochemistry and a Center for Applied Research and Rural Innovation grant at SFASU.

**Data Availability Statement:** Data are available from the corresponding author (PP) on personal request.

**Acknowledgments:** BS acknowledges support from the Welch Foundation Grant (No. AN-0008) at the Department of Chemistry and Biochemistry, and a Center for Applied Research and Rural Innovation grant at Stephen F. Austin State University, Nacogdoches, Texas, USA; PP acknowledges partial support from the National Institute of Health (Grant No. R21CA260147) and HW acknowledges support from National Institute of Health (Grant No. R03DE032766).

**Appendix A:** Characterization of the Aptamer-templated Ag-NC using the absorption and fluorescence spectroscopy:

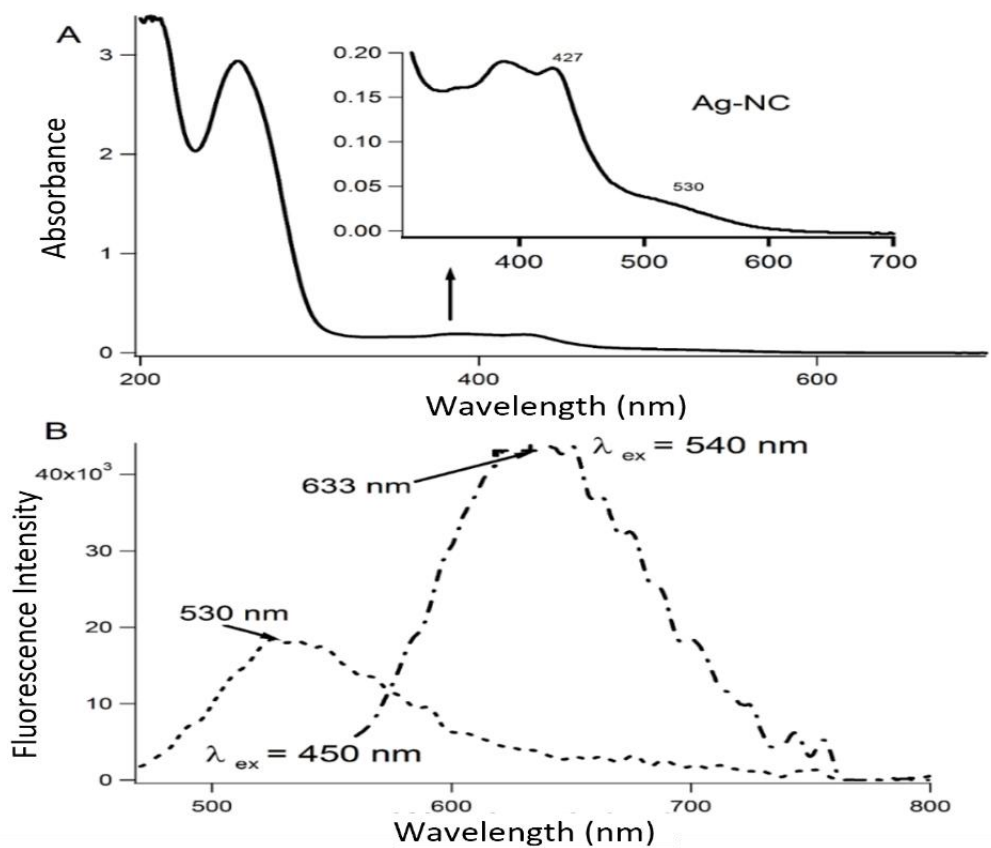

**Figure A1** (A and B) shows typical absorption and fluorescence spectra of aptamer-DNA templated Ag-NC. The inset in Figure A focuses on the Ag-NC's absorption region only.

The absorption and fluorescence spectroscopic characterization of the Ag-NCs in aptamer has been reported in our earlier publication using a series of DNA aptamers [8]. We re-experimented the most efficient targeted aptamer [8] for forming Ag-NC and verifying the repeatability of our previous work. The results show similar absorption and fluorescence emission profile features to our earlier publication [8], indicating Ag-NC formation on the aptamer template.

Figure A1 shows typical absorption and fluorescence emission spectra of aptamer-DNA 5'-CCC CCG TTG CTT TCG CTT TTC CTT TCG CTT TTG TTC GTT TCG TCC CTG CTT CCT TTC TTG-3' templated Ag-NC. The Ag-NCs were made in deionized water with DNA (15 µM), AgNO$_3$ (90 µM), and NaBH$_4$ (90 µM) to optimize nanoparticle formation, which is shown by the absorption band around the 380-390 nm region. The absorption of the nucleobases in the UV region is shown in Figure A1, while the inset highlights the absorbance of Ag-NC. An absorption band with a peak around 427 nm and a shorter band of ~ 530 nm were observed, proving the formation of more than one type of silver nanoclusters on this PA-specific aptamer template, which agrees with our previous work [8]. The emission profiles of these Ag-NCs show that for $\lambda_{ex}$ = 540 nm, the NC showed stronger fluorescence with $\lambda_{em}^{max}$ at ~ 633 nm compared to the emission band with $\lambda_{em}^{max}$ at 530 nm for $\lambda_{ex}$ = 450 nm, agreeing with our earlier work [8].

## References

1. F. Cohn, *Beiträge Zur Biologie Der Bacillen*, 7 (Beiträge zur Biologie der Pflanzen. Untersuchungen über Bacterien, 1877), Vol. IV, pp. 249–276.

2. H. Vlamakis, Y. Chai, P. Beauregard, R. Losick, and R. Kolter, "Sticking together: building a biofilm the Bacillus subtilis way," Nat. Rev. Microbiol. **11**, 157–168 (2013).

3. A. V. Samrot, A. Abubakar Mohamed, E. Faradjeva, L. Si Jie, C. Hooi Sze, A. Arif, T. Chuan Sean, E. Norbert Michael, C. Yeok Mun, N. Xiao Qi, P. Ling Mok, and S. S. Kumar, "Mechanisms and Impact of Biofilms and Targeting of Biofilms Using Bioactive Compounds—A Review," Medicina (Mex.) **57**, 839 (2021).

4. S. Auger, N. Ramarao, C. Faille, A. Fouet, S. Aymerich, and M. Gohar, "Biofilm Formation and Cell Surface Properties among Pathogenic and Nonpathogenic Strains of the Bacillus cereus Group," Appl. Environ. Microbiol. **75**, 6616–6618 (2009)

5. M. Shemesh and I. Ostrov, "Role of *Bacillus* species in biofilm persistence and emerging antibiofilm strategies in the dairy industry," J. Sci. Food Agric. **100**, 2327–2336 (2020).

6. M. Lemos, A. Borges, J. Teodósio, P. Araújo, F. Mergulhão, L. Melo, and M. Simões, "The effects of ferulic and salicylic acids on Bacillus cereus and Pseudomonas fluorescens single- and dual-species biofilms," Int. Biodeterior. Biodegrad. **86**, 42–51 (2014).

7. B. Sengupta, S. S. Sinha, B. L. Garner, I. Arany, C. Corley, K. Cobb, E. Brown, and P. C. Ray, "Influence of Aptamer-Enclosed Silver Nanocluster on the Prevention of Biofilm by Bacillus thuringiensis," Nanosci. Nanotechnol. Lett. **8**, 1054–1060 (2016).

8. B. Sengupta, P. Adhikari, E. Mallet, R. Havner, and P. Pradhan, "Spectroscopic Study on Pseudomonas Aeruginosa Biofilm in the Presence of the Aptamer-DNA Scaffolded Silver Nanoclusters," Molecules **25**, 3631 (2020).

9. M. T. T. Thi, D. Wibowo, and B. H. A. Rehm, "Pseudomonas aeruginosa Biofilms," Int. J. Mol. Sci. **21**, 8671 (2020).

10. G. Ramage and C. Williams, "The Clinical Importance of Fungal Biofilms," in *Advances in Applied Microbiology* (Elsevier, 2013), Vol. 84, pp. 27–83.

11. T. V. M. Vila and S. Rozental, "Biofilm Formation as a Pathogenicity Factor of Medically Important Fungi," in *Fungal Pathogenicity*, S. Sultan, ed. (InTech, 2016).

12. S. Fanning and A. P. Mitchell, "Fungal Biofilms," PLOS Pathog. **8**, e1002585 (2012).

13. Z. Malinovská, E. Čonková, and P. Váczi, "Biofilm Formation in Medically Important Candida Species," J. Fungi **9**, 955 (2023).

14. D. Davies, "Understanding biofilm resistance to antibacterial agents," Nat. Rev. Drug Discov. **2**, 114–122 (2003).

15. J. Monte, A. C. Abreu, A. Borges, L. C. Simões, and M. Simões, "Antimicrobial Activity of Selected Phytochemicals against Escherichia coli and Staphylococcus aureus and Their Biofilms," Pathogens **3**, 473–498 (2014).

16. V. S. Gondil and B. Subhadra, "Biofilms and their role on diseases," BMC Microbiol. **23**, 203, s12866-023-02954–2 (2023).

17. P. S. Stewart and T. Bjarnsholt, "Risk factors for chronic biofilm-related infection associated with implanted medical devices," Clin. Microbiol. Infect. **26**, 1034–1038 (2020).

18. G. M. Abebe, "The Role of Bacterial Biofilm in Antibiotic Resistance and Food Contamination," Int. J. Microbiol. **2020**, 1–10 (2020).

19. P. Lens, V. O'Flaherty, A. Moran, P. Stoodley, and T. Mahony, "Biofilms in Medicine, Industry and Environmental Biotechnology - Characteristics, Analysis and Control," Water Intell. Online **6**, 9781780402161–9781780402161 (2015).

20. A. Mishra, A. Aggarwal, and F. Khan, "Medical Device-Associated Infections Caused by Biofilm-Forming Microbial Pathogens and Controlling Strategies," Antibiotics **13**, 623 (2024).

21. M. Jamal, W. Ahmad, S. Andleeb, F. Jalil, M. Imran, M. A. Nawaz, T. Hussain, M. Ali, M. Rafiq, and M. A. Kamil, "Bacterial biofilm and associated infections," J. Chin. Med. Assoc. **81**, 7–11 (2018).

22. E. Hernández-Jiménez, R. Del Campo, V. Toledano, M. T. Vallejo-Cremades, A. Muñoz, C. Largo, F. Arnalich, F. García-Rio, C. Cubillos-Zapata, and E. López-Collazo, "Biofilm vs. planktonic bacterial mode of growth: Which do human macrophages prefer?," Biochem. Biophys. Res. Commun. **441**, 947–952 (2013).

23. "The use of superparamagnetic nanoparticles for prosthetic biofilm prevention," https://www.tandfonline.com/doi/epdf/10.2147/IJN.S5976?needAccess=true&role=button.

24. W. Wu, Z. Wu, T. Yu, C. Jiang, and W.-S. Kim, "Recent progress on magnetic iron oxide nanoparticles: synthesis, surface functional strategies and biomedical applications," Sci. Technol. Adv. Mater. **16**, 023501 (2015).

25. M. P. Ferraz, "Advanced Nanotechnological Approaches for Biofilm Prevention and Control," Appl. Sci. **14**, 8137 (2024).

26. P. C. Balaure and A. M. Grumezescu, "Recent Advances in Surface Nanoengineering for Biofilm Prevention and Control. Part I: Molecular Basis of Biofilm Recalcitrance. Passive Anti-Biofouling Nanocoatings," Nanomaterials **10**, 1230 (2020).

27. S. Gurunathan, J. W. Han, D.-N. Kwon, and J.-H. Kim, "Enhanced antibacterial and anti-biofilm activities of silver nanoparticles against Gram-negative and Gram-positive bacteria," Nanoscale Res. Lett. **9**, 373 (2014).

28. A. S. Joshi, P. Singh, and I. Mijakovic, "Interactions of Gold and Silver Nanoparticles with Bacterial Biofilms: Molecular Interactions behind Inhibition and Resistance," Int. J. Mol. Sci. **21**, 7658 (2020).

29. D. Żyro, J. Sikora, M. I. Szynkowska-Jóźwik, and J. Ochocki, "Silver, Its Salts and Application in Medicine and Pharmacy," Int. J. Mol. Sci. **24**, 15723 (2023).

30. Y. K. Mohanta, K. Biswas, S. K. Jena, A. Hashem, E. F. Abd_Allah, and T. K. Mohanta, "Anti-biofilm and Antibacterial Activities of Silver Nanoparticles Synthesized by the Reducing Activity of Phytoconstituents Present in the Indian Medicinal Plants," Front. Microbiol. **11**, 1143 (2020).

31. B. Hosnedlova, D. Kabanov, M. Kepinska, V. H. B Narayanan, A. A. Parikesit, C. Fernandez, G. Bjørklund, H. V. Nguyen, A. Farid, J. Sochor, A. Pholosi, M. Baron, M. Jakubek, and R. Kizek, "Effect of Biosynthesized Silver Nanoparticles on Bacterial Biofilm Changes in S. aureus and E. coli," Nanomaterials **12**, 2183 (2022).

32. P. R. More, S. Pandit, A. D. Filippis, G. Franci, I. Mijakovic, and M. Galdiero, "Silver Nanoparticles: Bactericidal and Mechanistic Approach against Drug Resistant Pathogens," Microorganisms **11**, 369 (2023).

33. J. G. Bruno and M. P. Carrillo, "Development of Aptamer Beacons for Rapid Presumptive Detection of Bacillus Spores," J. Fluoresc. **22**, 915–924 (2012).

34. M. Ikanovic, W. E. Rudzinski, J. G. Bruno, A. Allman, M. P. Carrillo, S. Dwarakanath, S. Bhahdigadi, P. Rao, J. L. Kiel, and C. J. Andrews, "Fluorescence Assay Based on Aptamer-Quantum Dot Binding to Bacillus thuringiensis Spores," J. Fluoresc. **17**, 193–199 (2007).

35. M. Domsicova, J. Korcekova, A. Poturnayova, and A. Breier, "New Insights into Aptamers: An Alternative to Antibodies in the Detection of Molecular Biomarkers," Int. J. Mol. Sci. **25**, 6833 (2024).

36. H. R. Im, S. J. Im, D. V. Nguyen, S. P. Jeong, and A. Jang, "Real-time diagnosis and monitoring of biofilm and corrosion layer formation on different water pipe materials using non-invasive imaging methods," Chemosphere **361**, 142577 (2024).

37. K. He, X. Zhang, S. Ren, and J. Sun, "Deep Residual Learning for Image Recognition," in (2016), pp. 770–778.

38. M. Drozdzal, E. Vorontsov, G. Chartrand, S. Kadoury, and C. Pal, "The Importance of Skip Connections in Biomedical Image Segmentation," in *Deep Learning and Data Labeling for Medical Applications*, G. Carneiro, D. Mateus, L. Peter, A. Bradley, J. M. R. S. Tavares, V. Belagiannis, J. P. Papa, J. C. Nascimento, M. Loog, Z. Lu, J. S. Cardoso, and J. Cornebise, eds., Lecture Notes in Computer Science (Springer International Publishing, 2016), Vol. 10008, pp. 179–187.

39. G. Dimauro, F. Deperte, R. Maglietta, M. Bove, F. La Gioia, V. Renò, L. Simone, and M. Gelardi, "A novel approach for biofilm detection based on a convolutional neural network," Electronics **9**, 881 (2020).

40. O. Ronneberger, P. Fischer, and T. Brox, "U-Net: Convolutional Networks for Biomedical Image Segmentation," in *Medical Image Computing and Computer-Assisted Intervention – MICCAI 2015*, N. Navab, J. Hornegger, W. M. Wells, and A. F. Frangi, eds., Lecture Notes in Computer Science (Springer International Publishing, 2015), pp. 234–241.

41. J. Le'Clerc Arrastia, N. Heilenkötter, D. Otero Baguer, L. Hauberg-Lotte, T. Boskamp, S. Hetzer, N. Duschner, J. Schaller, and P. Maass, "Deeply Supervised UNet for Semantic Segmentation to Assist Dermatopathological Assessment of Basal Cell Carcinoma," J. Imaging **7**, 71 (2021).

42. R. Das, A. Dhiman, A. Kapil, V. Bansal, and T. K. Sharma, "Aptamer-mediated colorimetric and electrochemical detection of Pseudomonas aeruginosa utilizing peroxidase-mimic activity of gold NanoZyme," Anal. Bioanal. Chem. **411**, 1229–1238 (2019).

43. B. Sengupta, C. M. Ritchie, J. G. Buckman, K. R. Johnsen, P. M. Goodwin, and J. T. Petty, "Base-Directed Formation of Fluorescent Silver Clusters," J. Phys. Chem. C **112**, 18776–18782 (2008).

44. E. Haney, M. Trimble, J. Cheng, Q. Vallé, and R. Hancock, "Critical Assessment of Methods to Quantify Biofilm Growth and Evaluate Antibiofilm Activity of Host Defence Peptides," Biomolecules **8**, 29 (2018).

45. U. Tatli and C. Budak, "Biomedical Image Segmentation with Modified U-Net," Trait. Signal **40**, 523–531 (2023).

46. N. Otsu, "A Threshold Selection Method from Gray-Level Histograms," IEEE Trans. Syst. Man Cybern. **9**, 62–66 (1979).